\documentclass[aps, prl, reprint, superscriptaddress, longbibliography]{revtex4-2}

\usepackage[utf8]{inputenc}
\usepackage[export]{adjustbox}
\usepackage{tikz}
\usepackage{amsmath}
\usepackage{amssymb}
\usepackage{hyperref}
\usepackage{graphicx}
\usepackage{epstopdf}
\usepackage{dcolumn}
\usepackage{soul}
\usepackage{mathrsfs}
\usepackage{bm}
\usepackage{soul}
\usepackage{xspace}
\usepackage{verbatim}
\usepackage{color}
\usepackage{xcolor}

\hypersetup{colorlinks=true,linkcolor=blue,citecolor=blue, filecolor=blue,urlcolor=blue,breaklinks=true}

\begin{document}

\title{Extractable Information Capacity in Sequential Measurements Metrology}

\author{Yaoling Yang}
\email{yyaoling@std.uestc.edu.cn}
\affiliation{Institute of Fundamental and Frontier Sciences, University of Electronic Science and Technology of China, Chengdu 610051, China}

\author{Victor Montenegro}
\email{vmontenegro@uestc.edu.cn}
\affiliation{Institute of Fundamental and Frontier Sciences, University of Electronic Science and Technology of China, Chengdu 610051, China}

\author{Abolfazl Bayat}
\email{abolfazl.bayat@uestc.edu.cn}
\affiliation{Institute of Fundamental and Frontier Sciences, University of Electronic Science and Technology of China, Chengdu 610051, China}

\date{\today}

\begin{abstract}
The conventional formulation of quantum sensing is based on the assumption that the probe is reset to its initial state after each measurement. In a very distinct approach, one can also pursue a sequential measurement scheme in which time-consuming resetting is avoided. In this situation, every measurement outcome effectively comes from a different probe, yet correlated with other data samples. Finding a proper description for the precision of sequential measurement sensing is very challenging as it requires the analysis of long sequences with exponentially large outcomes. Here, we develop a recursive formula and an efficient Monte-Carlo approach to calculate the Fisher information, as a figure of merit for sensing precision, for arbitrary lengths of sequential measurements. Our results show that Fisher information initially scales non-linearly with the number of measurements and then asymptotically saturates to linear scaling. Such transition, which fundamentally constrains the extractable information about the parameter of interest, is directly linked to the finite memory of the probe when undergoes multiple sequential measurements. Based on these, we establish a figure of merit to determine the optimal measurement sequence length and exemplify our results in three different physical systems.
\end{abstract}

\maketitle

\textit{Introduction.---} Quantum probes exhibit unparalleled precision compared with their classical counterparts for a given resource~\cite{Degen}. The resource efficiency of quantum probes has been demonstrated through exploiting the superposition principle via Greenberger-Horne-Zeilinger-type entanglement~\cite{Boto-2000,Leibfried-2004,Giovannetti2004, Giovannetti2006,Giovannetti2011,Frowis2011,Demkowicz2012}, criticality in many-body systems~\cite{Ding2022, Liu2021, zanardi2008quantum, invernizzi2008optimal, salvatori2014quantum, zanardi2007critical, Garbe2020, Jianming2021critical, Horodecki2018prx, Sarkar2022topological, PhysRevX.8.021022, PhysRevLett.126.200501}, variational methods~\cite{meyer2021variational,marciniak2021optimal,yang2021variational}, adaptive~\cite{higgins2007entanglement,Said,Berry2009,Higgins2009,Bonato2015} or continuous measurements~\cite{PRXQuantum.3.010354, gammelmark2014fisher,PhysRevLett.125.200505, Albarelli_2017-continuous, Albarelli2018restoringheisenberg, cabot2022quantum, cabot2023continuous}, and Floquet dynamics~\cite{Mishra2021, Mishra2022}, to name a few. In a general quantum sensing scenario, for estimating an unknown parameter $\lambda$, encoded in the quantum state $\rho_\lambda$, one performs a measurement in a certain basis and then feeds the outcomes to a classifier. The precision is bounded through Cram\'{e}r-Rao inequality $\delta \lambda{\geq}1{/}\sqrt{M\mathcal{F}_\lambda}$, where $\delta\lambda$ is the uncertainty in estimating $\lambda$, $M$ is the number of trials, and $\mathcal{F}_\lambda$ is the Fisher Information (FI)~\cite{Holevo, cramer1999mathematical, LeCam-1986, helstrom1969quantum,zamir1998proof}. While classical sensors exhibit linear scaling of the FI with respect to a given resource (e.g., measurement time), quantum probes demonstrate greater resource efficiency, enabling the possibility of achieving super-linear scaling. The formulation of the Cram\'{e}r-Rao inequality assumes the resetting of the probe after each measurement or equivalently using $M$ identical probes at once. By avoiding the time-consuming resetting procedure, the quantum state of the probe would be different for each data sample, implying the use of several non-identical probes for sensing an unknown parameter. Therefore, a question arises: how will the precision scale when measurements are performed sequentially without resetting the probe?

In many-body systems with partial accessibility, local measurements on a subsystem lead to a global wave function collapse, which has been the subject of intensive studies~\cite{Busch1990, Schmidt_2020, BAN2021127383, PhysRevX.9.031009, PhysRevLett.128.010604, benoist2023limit, Haapasalo_2016, benoist2019invariant, burgarth2014exponential, pouyandeh2014measurement, Bayat, ma2018phase,PhysRevA.92.042315}. In the domain of quantum metrology, sequences of projective measurements followed by free evolution at regular time intervals have led to Hamiltonian identification~\cite{HMabuchi_1996}, sequential measurements sensing schemes~\cite{Burgarth2015, PhysRevA.96.012316, Ritboon_2022, bompais2023asymptotic, 9992617, proceedings2019012011, PhysRevA.64.042105, PhysRevA.94.042322, Nagali_2012, PhysRevA.92.032124}, and even hinting quantum-enhanced sensitivity observed in short sequences that are computationally feasible~\cite{PhysRevA.99.022102, PhysRevLett.129.120503}. Studying sequential measurements sensing schemes is typically limited to short lengths of measurement sequences. The limitation arises from the exponential growth of measurement outcomes with the number of measurement sequences. Indirect approaches, based on a functional of the measurement outcomes~\cite{Burgarth2015} or correlated stochastic processes~\cite{Radaelli_2023}, have been proposed to estimate sensing precision with a large number of sequential measurements. Interestingly, for short (${\sim}20$) measurement sequences, it has been demonstrated that the FI scales super-linearly with the number of measurements~~\cite{PhysRevA.99.022102, PhysRevLett.129.120503}, whereas, the FI stemmed from indirect methods scales linearly in the asymptotic limit of sequential measurements~\cite{Burgarth2015}. To reconcile this apparent discrepancy, an efficient approach for studying the FI for an arbitrary number of measurements is highly desirable. 

In this Letter, we develop a recursive formula and an efficient Monte-Carlo approach to compute the FI for arbitrary lengths of measurement sequences. Our results show that the FI exhibits non-linear growth in the beginning and asymptotically saturates to a linear function. Interestingly, this transition directly relates to the memory of an early state. In fact, the quantum probe retains only a finite memory of its initial state during multiple sequences of evolution and local projective measurements. Once the probe loses its memory, the FI grows linearly with additional measurements. This enables us to establish a figure of merit for determining the optimal reset point, maximizing the sensing protocol. To support our findings, we investigate three distinct physical systems in both closed and open quantum systems.

\textit{Sequential measurement sensing protocol.---} Conventional sensing schemes typically rely on measurement outcomes with independent and identically distributed (i.i.d.) probability distributions. Hence, after measuring the probe, it is necessary to reset the sensing procedure to its exact initial quantum state in preparation for another round of measurements. These requirements may result in resource-demanding state preparation and time overhead due to unavoidable resetting procedures. On the other hand, sequential sensing schemes~\cite{PhysRevLett.129.120503,PhysRevA.99.022102, Burgarth2015} utilize non-i.i.d. probability distributions constructed from consecutive measurements on the probe at regular intervals. Let us consider a probe initialized in a quantum state $\rho^{(1)}(0){=}\rho_0$. The sequential sensing protocol is an iterative approach: (i) the quantum probe $\rho^{(i)}(0)$ freely evolves to $\rho^{(i)}(\tau_i){=}U^{(i)}_\lambda\rho^{(i)}(0)U_\lambda^{(i)\dagger}$ with a unitary time evolution operator $U_\lambda^{(i)}$; (ii) at time $\tau_i$ a local positive operator-valued measure (POVM) $\{\Pi_{\gamma_i}\}$ with random outcome $\gamma_i$ is performed on the probe, collapsing the state into $\rho^{(i{+}1)}(0){=}\Pi_{\gamma_i}\rho^{(i)}(\tau_i)\Pi_{\gamma_i}^\dagger/p(\gamma_i)$, where $p(\gamma_i){=}\text{Tr}[\Pi_{\gamma_i}\rho^{(i)}(\tau_{i})\Pi_{\gamma_i}^\dagger]$ is the probability associated to $\gamma_i$ at step $i$; (iii) the outcome $\gamma_i$ is recorded and the new initial state $\rho^{(i{+}1)}(0)$ is replaced in (i); (iv) the above steps are repeated until $n_\mathrm{seq}$ measurements outcomes are consecutively obtained; (v) after gathering a data sequence $\pmb{\gamma}{=}(\gamma_1,{\cdots},\gamma_{n_\mathrm{seq}})$, the probe is reset to $\rho_0$ and the process is repeated to generate a new trajectory.

\textit{Fisher information.---} The FI is given by:
\begin{equation}
\mathcal{F}_\lambda=\sum_{\pmb{\gamma}} P_{\pmb{\gamma}} \left(\partial_\lambda \ln P_{\pmb{\gamma}}\right)^{2}, \hspace{1cm} P_{\pmb{\gamma}}=\prod_{\gamma_i=1}^{n_{\mathrm{seq}}}p(\gamma_i),\label{eq:CFI_trajectories}
\end{equation}
where $\sum_{\pmb{\gamma}}$ runs over all possible trajectories, $\partial_\lambda{:=}\partial{/}\partial \lambda$, and $P_{\pmb{\gamma}}$ is the conditional probability associated with a particular quantum trajectory $\pmb{\gamma}$. Note that the above FI quantifies the achievable precision limit for a given measurement basis, considering all trajectories $\pmb{\gamma}$. However, the exponential growth in the number of trajectories $\pmb{\gamma}$ makes the computation of the FI in Eq.~\eqref{eq:CFI_trajectories} infeasible. We now address this issue with an efficient method for arbitrary numbers of sequential measurements.

\textit{Sequential-based metrology for long trajectories.---} The recursive formula of the FI for arbitrary $n_\mathrm{seq}$ follows:

\noindent \textbf{Proposition:} The information gained about $\lambda$ after performing a subsequent $n$ measurement on the probe conditioned on all previous $n{-}1$ measurements is:
\begin{equation}
\mathcal{F}_\lambda^{(n)}{=}\mathcal{F}_\lambda^{(n{-}1)}{+}{\Delta}\mathcal{F}_\lambda^{(n)};\hspace{0.5cm}\Delta\mathcal{F}_\lambda^{(n)}{:=}\sum_{\pmb{\gamma}} P_{\pmb{\gamma}} f_\lambda^{\pmb{\gamma},(n)},\label{eq_joint-cfi-us}
\end{equation}
where $\mathcal{F}_\lambda^{(n)}$ is the FI at step $n$, $\Delta\mathcal{F}_\lambda^{(n)}$ is the increment of the FI after performing one more measurement while the previous $n{-}1$ outcomes have been recorded, and $f_\lambda^{\pmb{\gamma},(n)}$ is the FI obtained from the $n$-th measurement $p(\gamma_n)$ in trajectory $\pmb{\gamma}$.

\noindent \textit{Proof:} From conditional probabilities, one gets:
\begin{equation}
    \ln P_{\pmb{\gamma}^{(n)}} = \ln P_{\pmb{\gamma}^{(n{-}1)}}+\ln p(\gamma_{n}{\mid}\pmb{\gamma}^{(n{-}1)}),\label{eq_conditional_prob}
\end{equation}
where $\pmb{\gamma}^{(n)}$ is a generic trajectory with $n$ measurement outcomes, $p(\gamma_{n}{\mid}\pmb{\gamma}^{(n{-}1)})$ is the conditional probability of obtaining the outcome $\gamma_n$ at step $n$ conditioned upon all $n{-}1$ previous measurement outcomes. Replacing Eq.~\eqref{eq_conditional_prob} into Eq.~\eqref{eq:CFI_trajectories},  see Supplemental Material (SM)~\cite{SM} for details, results in:
\begin{equation}
\mathcal{F}_\lambda^{(n)}{=}\mathcal{F}_\lambda^{(n{-}1)}{+}{\sum_{\pmb{\gamma}^{(n{-}1)}}}P_{\pmb{\gamma}^{(n{-}1)}} {\sum_{\gamma_{n}}}\frac{\left[\partial_\lambda p(\gamma_{n}{\mid} \pmb{\gamma}^{(n{-}1)})\right]^{2}}{p(\gamma_{n}{\mid}\pmb{\gamma}^{(n{-}1)})},
\end{equation}
which leads to $\mathcal{F}_\lambda^{(n)}=\mathcal{F}_\lambda^{(n{-}1)}{+}\sum_{\pmb{\gamma}} P_{\pmb{\gamma}}f_\lambda^{\pmb{\gamma},(n)} \blacksquare$.

Note that Eq.~\eqref{eq_joint-cfi-us} is computationally infeasible as it considers the whole trajectory space $\pmb{\gamma}$. To circumvent this, we employ a Monte-Carlo approach to evaluate the FI for any $n_\mathrm{seq}$ (see SM~\cite{SM} for numerical robustness), approximating the FI increment as follows:
\begin{equation}
\Delta \mathcal{F}_\lambda^{(n)} \sim \sum_{\mu{=}1}^{\mu_{\mathrm{max}}} \frac{f_\lambda^{\mu, (n)}}{\mu_{\mathrm{max}}}, \label{eq:mc_delta_f}
\end{equation}
where $f_\lambda^{\mu,{(n)}}$ is the FI obtained from $p(\gamma_n)$ in Monte-Carlo trajectory $\mu$, and $\mu_{max}$ is the total numbers of Monte-Carlo sampling. In the limit of $\mu_\mathrm{max}{\rightarrow}\infty$, Eq.~\eqref{eq:mc_delta_f} converges to the actual FI increment shown in Eq.~\eqref{eq_joint-cfi-us}. One key question arises from Eq.~\eqref{eq:mc_delta_f}: how does the increment $\Delta\mathcal{F}_\lambda^{(n)}$ of FI behave as $n$ increases? Is there a limit to the amount of information that can be extracted about $\lambda$ as $n$ increases? To address these issues, we shed light on two central features of sequential measurements sensing: memory loss and stationary steady state.

\begin{figure}[t]
\includegraphics[width=\linewidth]{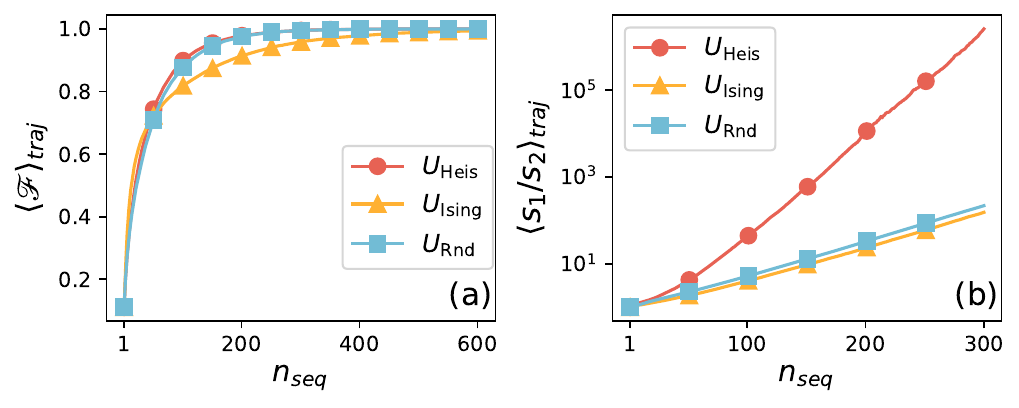} 
\caption{We compare final states $\phi^{(n_\mathrm{seq})}$ and $\theta^{(n_\mathrm{seq})}$ resulting from two distinct random initial states $\rho_0{=}\phi_0$ and $\rho_0{=}\theta_0$ following the \emph{same} quantum trajectory. (a) Fidelity 
$\langle\mathscr{F}\rangle_\mathrm{traj}{:=}\langle\mathscr{F}(\phi^{(n_\mathrm{seq})}{,}\theta^{(n_\mathrm{seq})})\rangle_\mathrm{traj}$ as a function of $n_\mathrm{seq}$ for several unitary operators. (b) Ratio between the first two largest singular values $s_1$ and $s_2$ as a function of $n_\mathrm{seq}$ for several unitary operators. We consider a system size of $N{=}6$.} \label{fig_fidelity} 
\end{figure}

\textit{Memory loss.---} Performing $n_\mathrm{seq}$ sequential measurements on the probe's initial state $\rho_0$ leads to a gradual loss of information about $\rho_0$. To support this, consider $V^{(j)}{:=}\Pi_{\gamma_j}U_\lambda^{(j)}$, where $U^{(j)}$ is a unitary operator and $\Pi_{\gamma_j}{=}\mathbb{I}{\otimes}|\gamma_j\rangle\langle\gamma_j|$ is a local projection measurement. The (unnormalized) quantum state after $n_\mathrm{seq}$ steps is $\rho^{(n_\mathrm{seq})}{\sim}\mathscr{P}_\lambda^{(n_\mathrm{seq})}\rho_0 \mathscr{P}_\lambda^{(n_\mathrm{seq})\dagger}$, where $\mathscr{P}_\lambda^{(n_\mathrm{seq})}{:=}\prod_{j=1}^{n_\mathrm{seq}} V^{(j)}$. To validate the memory loss observation, we simulate final states $\phi^{(n_\mathrm{seq})}$ and $\theta^{(n_\mathrm{seq})}$ resulting from two distinct random initial states $\rho_0{=}\phi_0$ and $\rho_0{=}\theta_0$ following the \emph{same} quantum trajectory. We consider three $\mathscr{P}_\lambda^{(n_\mathrm{seq})}$ cases: (i) a random unitary $U_\mathrm{Rnd}$ measured locally in computational basis; (ii) a Heisenberg unitary $U_\mathrm{Heis}{=}\exp[-i\tau H_\mathrm{Heis}]$, where $H_\mathrm{Heis}{=}{-}J\sum_{j=1}^{N-1}\boldsymbol{\sigma}_{j}{\cdot}\boldsymbol{\sigma}_{j+1}$; and (iii) an Ising unitary $U_\mathrm{Ising}{=}\exp[-i\tau H_\mathrm{Ising}]$, where $H_\mathrm{Ising}{=}{-}J \sum_{j=1}^{N-1}\sigma_{j}^z\sigma_{j+1}^z{+}B\sum_{j=1}^N\sigma_{j}^{x}$. In the above, $N$ is the system size, $\boldsymbol{\sigma}_{j}{=}(\sigma_j^x{,}\sigma_j^y{,}\sigma_j^z)$ is a vector of Pauli matrices acting at site $j$, $J{>}0$ is the exchange interaction, and $B$ is a magnetic field. For the Heisenberg (Ising) case we sequentially measured a single spin in $\sigma_z$ ($\sigma_x$) basis at $J\tau{=}N$. To quantify their distinguishability between states, we use the fidelity $\mathscr{F}{:=}\mathscr{F}(\phi^{(n_\mathrm{seq})}{,}\theta^{(n_\mathrm{seq})})$~\cite{nielsen00}. In Fig.~\ref{fig_fidelity}(a), we plot the fidelity $\langle\mathscr{F}\rangle_\mathrm{traj}$ averaged over $10^4$ trajectories as a function of $n_\mathrm{seq}$ for several unitary operators. As the figure shows, the fidelity goes towards unity as $n_\mathrm{seq}$ increases, namely: the resulting states are indistinguishable regardless of their initial states.

\textit{Stationary steady state.---} Interestingly, the total evolution-measurement operator $\mathscr{P}_\lambda^{(n_\mathrm{seq})}$ approximates a rank-1 matrix as $n_\mathrm{seq}$ increases. Hence, $\mathscr{P}_\lambda^{(n_\mathrm{seq})}$ can be written as $\mathscr{P}_\lambda^{(n_\mathrm{seq})}{\sim}|p\rangle\langle p|$, where $|p\rangle$ is an orthonormal basis of the singular value decomposition of $\mathscr{P}_\lambda^{(n_\mathrm{seq})}$. In Fig.~\ref{fig_fidelity}(b), we plot the ratio between the first two largest singular values of $\mathscr{P}_\lambda^{(n_\mathrm{seq})}$, namely $s_1$ and $s_2$, as a function of $n_\mathrm{seq}$ for different unitary operators. As seen from the figure, $s_1$ dominates as $n_\mathrm{seq}$ increases, resulting in a rank-1 matrix in agreement with the memory loss of the initial state, see Fig.~\ref{fig_fidelity}(a). Thus, for a given trajectory the state of the system asymptotically approaches its stationary steady state $|p\rangle\langle p|$ independent of its initial state.

\textit{Extractable information limits.---} We aim to estimate the parameter $\lambda$ which is  encoded in the unitary operation $U^{(j)}_\lambda$. Let us assume that $\mathscr{P}_\lambda^{(m)}$ becomes a rank-1 matrix after $m$ sequential measurements, that is $\mathscr{P}_\lambda^{(m)}{\sim}|p(m,\lambda)\rangle\langle p(m,\lambda)|$. This means that for any arbitrary long sequence, the final state can always be written as: $\rho^{(m)}{\sim}\mathscr{P}_\lambda^{(m)}\tilde{\rho}\mathscr{P}_\lambda^{\dagger(m)}$, where $\tilde{\rho}$ is any density matrix independent of $\lambda$. Since $m$ is a finite number of sequential steps, the FI that can be accumulated throughout this process can only be finite. The above statement implies:
\begin{equation}
\Delta\mathcal{F}_\lambda^{(j)}{=}\sum_{\pmb{\gamma}} P_{\pmb{\gamma}} f_\lambda^{\pmb{\gamma},(j)}{\leq}f_\lambda^{\text{max},(j)}{\sim}G(\mathscr{P}_\lambda^{(m)}),\label{eq_extractable_limits}
\end{equation}
where $f_\lambda^{\text{max},(j)}{=}\mathrm{max}_{\pmb{\gamma}}[f_\lambda^{\pmb{\gamma},(j)}]$ and $G(\mathscr{P}_\lambda^{(m)})$ is a finite function depending on $\mathscr{P}_\lambda^{(m)}$. Hence, at each measurement step one can, at best, add $G(\mathscr{P}_\lambda^{(m)})$ to the FI. The direct consequence is that the FI is bounded by a linear function of the number of sequential measurements $m$. In the following, we provide three distinct physical systems to support our findings.

\begin{figure}[t]
\includegraphics[width=\linewidth]{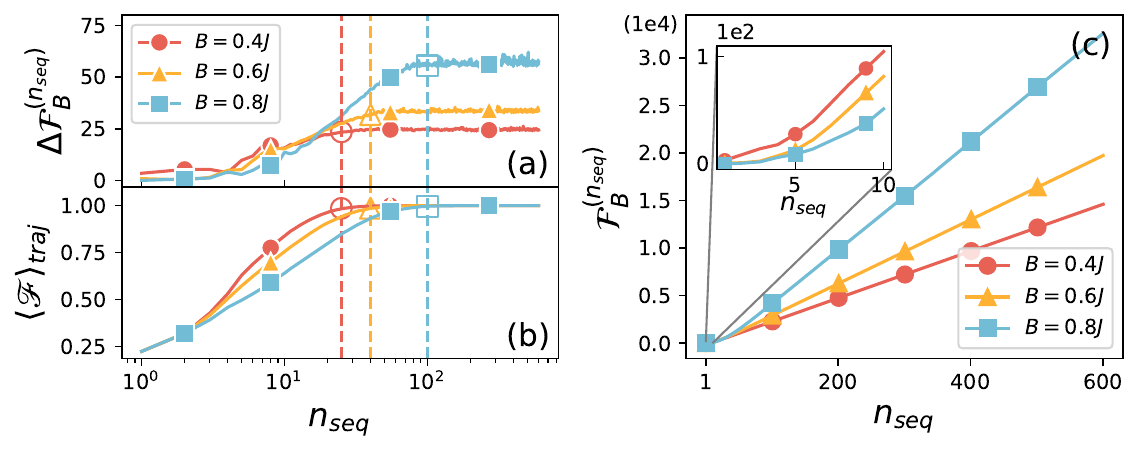} 
\caption{(a) FI increment $\Delta\mathcal{F}_{B}^{(n_\mathrm{seq})}$ as a function of $n_\mathrm{seq}$ for various $B$. (b) Averaged fidelity $\langle\mathscr{F}\rangle_\mathrm{traj}$ as a function of $n_\mathrm{seq}$ for various $B$. Dashed lines divide the curves into a non-trivial (memory) and constant (memoryless) dependence on $n_\mathrm{seq}$. (c) $\mathcal{F}_{B}^{(n_\mathrm{seq})}$ as a function of $n_\mathrm{seq}$ for several values of $B$.} \label{fig_delta_F_fidelity}
\end{figure}

\textit{Example 1: Spin chain magnetometry.---} We consider $N$ spin-$1{/}2$ particles with Heisenberg interaction in the presence of a local magnetic field $B$, which we aim to estimate with sequential metrology. The Hamiltonian is:
\begin{equation}
H=-J \sum_{j=1}^{N-1} \boldsymbol{\sigma}_{j} \cdot \boldsymbol{\sigma}_{j+1}+B \sigma_{1}^{x},\label{eq_spin_chain}
\end{equation} 
where $J{>}0$ is the exchange interaction and $B$ is a magnetic field. Without loss of generality, we consider a probe size $N{=}4$, where each trajectory initializes in $|\psi(0)\rangle{=}|\downarrow\rangle^{\otimes N}$, evolves unitarily at regular times $J\tau_i{=}J\tau{=}N$ under $U^{(j)}{=}U{=}e^{{-}i\tau H}$, and measured sequentially at local site $N$ in $\sigma_z$ basis. In Fig.~\ref{fig_delta_F_fidelity}(a), we plot the FI increment $\Delta\mathcal{F}_{B}^{(n_\mathrm{seq})}$ averaged over $10^5$ trajectories [see Eq.~\eqref{eq:mc_delta_f}] as a function of $n_\mathrm{seq}$ for several values of $B$. As the figure shows, $\Delta\mathcal{F}_{B}^{(n_\mathrm{seq})}$ initially grows with increasing $n_\mathrm{seq}$ and then saturates to an approximately constant value. The transition has been specified by a dashed line. To link the memory loss feature with the above non-trivial behaviour of $\Delta\mathcal{F}_{B}^{(n_\mathrm{seq})}$, we simulate final states $\phi^{(n_\mathrm{seq})}$ and $\theta^{(n_\mathrm{seq})}$ resulting from two distinct random initial states $\rho_0{=}\phi_0$ and $\rho_0{=}\theta_0$ following the \emph{same} quantum trajectory. In Fig.~\ref{fig_delta_F_fidelity}(b), we plot the fidelity $\langle\mathscr{F}\rangle_\mathrm{traj}$ averaged over $10^4$ trajectories as a function of $n_\mathrm{seq}$ for several values of $B$. Remarkably, as seen from the figure, a clear correspondence between the non-trivial dependence of $\Delta\mathcal{F}_{B}^{(n_\mathrm{seq})}$ with respect to $n_\mathrm{seq}$ and the loss of memory with respect to the probe's initial state emerges ---see dashed lines in Figs.~\ref{fig_delta_F_fidelity}(a)-(b). Indeed, as the probe keeps the memory of the initial state, the FI increment grows non-linearly with respect to $n_\mathrm{seq}$. Conversely, when the probe loses its memory of the initial state, the FI increment reaches an approximate constant value. Thus, the FI can only grow linearly with $n_\mathrm{seq}$. This is explicitly depicted in Fig.~\ref{fig_delta_F_fidelity}(c), where we plot the FI $\mathcal{F}_{B}^{(n_\mathrm{seq})}$ as a function of $n_\mathrm{seq}$ for several values of $B$. As shown in the figure, a clear super-linear behavior (inset) transits to a linear behaviour as $n_\mathrm{seq}$ increases.

\begin{figure}[t]
\includegraphics[width=\linewidth]{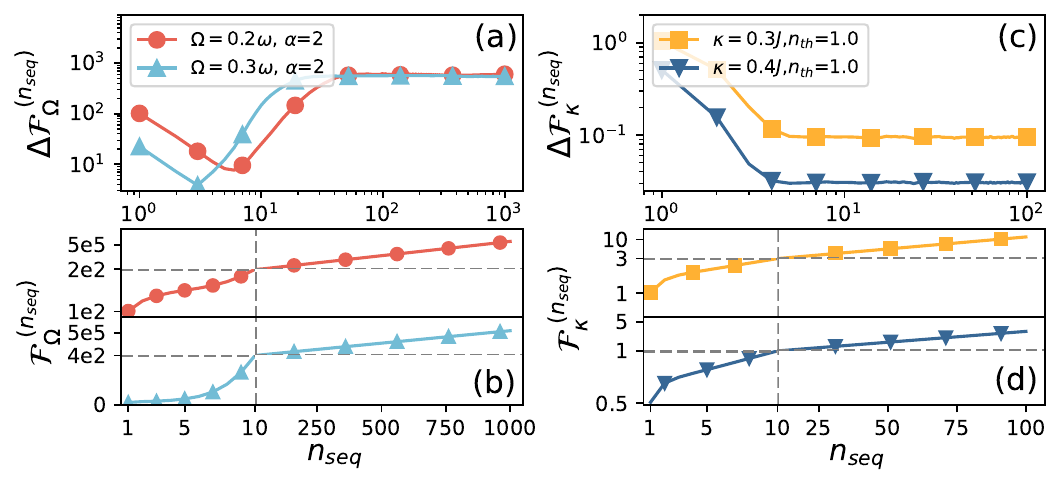} 
\caption{JC model case: (a) $\Delta\mathcal{F}_\Omega^{(n_\mathrm{seq})}$ as a function of $n_\mathrm{seq}$ for various $\Omega$. (b) $\mathcal{F}_\Omega^{(n_\mathrm{seq})}$ as a function of $n_\mathrm{seq}$ for several $\Omega$. Non-unitary case: (c) $\Delta\mathcal{F}_\kappa^{(n_\mathrm{seq})}$ as a function of $n_\mathrm{seq}$ for different values of $\kappa$, (d) $\mathcal{F}_\kappa^{(n_\mathrm{seq})}$ as a function of $n_\mathrm{seq}$ for various $\kappa$.}\label{fig_JC_lindblad} 
\end{figure}

\textit{Example 2: Light-matter interaction.---} We consider the Jaynes-Cummings (JC) model which describes the interaction between a two-level atom with a quantized radiation field~\cite{jcmodel}. The Hamiltonian is:
\begin{equation}
H_\mathrm{JC}{=}\hbar\omega_c a^\dagger a{+}\frac{1}{2}\hbar\omega_a \sigma^z{+}\hbar\Omega(\sigma^+a{+}\sigma^-a^\dagger),\label{eq_JC_model}
\end{equation}
where $a{(}a^\dagger{)}$ is the annihilation (creation) operator, $\sigma^z{=}|e\rangle\langle e|{-}|g\rangle\langle g|{,}\sigma^{+}{=}|e\rangle\langle g|{,}\sigma^{-}{=}|g\rangle\langle e|$, where $|g\rangle{(}|e\rangle{)}$ is the ground (excited) state of the two-level atom, $\omega_c$ is the frequency of the field, $\omega_a$ is the two-level atom's transition frequency, and $\Omega$ is the atom-field coupling strength. We aim to estimate $\Omega$ using sequential measurements on the atom. Without loss of generality, $\omega_c{=}\omega_a{=}\omega$, we initialize each trajectory from $|\psi(0)\rangle{=}|g\rangle|\alpha\rangle$, where $|\alpha\rangle$ is a coherent state, and measurements are performed at $\omega \tau{=}2\pi$ intervals. Notably, for $n_\mathrm{seq}{\gg}1$, the field will likely to be filtered into a specific number state $|\tilde{m}\rangle$ (see SM~\cite{SM} for details). This implies that the atom-field state evolves within the subspace $\{|e,\tilde{m}\rangle{,}|g,\tilde{m}{+}1\rangle\}$, with $\Delta\mathcal{F}_\Omega^{(n_\mathrm{seq})}{\sim}\tau^2(\tilde{m}{+}1)$. Hence, for a fixed evolution time $\tau$ and number state $\tilde{m}$, the FI increment is bounded in agreement with Eq.~\eqref{eq_extractable_limits}. In Fig.~\ref{fig_JC_lindblad}(a), we plot the FI increment $\Delta\mathcal{F}_\Omega^{(n_\mathrm{seq})}$ as a function of $n_\mathrm{seq}$ for two coupling strengths $\Omega$. As the figure shows, the transition between non-trivial to a constant value of $\Delta\mathcal{F}_\Omega^{(n_\mathrm{seq})}$ concerning $n_\mathrm{seq}$ holds. Thus, the FI can only grow linearly with extra measurements. In Fig.~\ref{fig_JC_lindblad}(b), we plot the FI $\mathcal{F}_{\Omega}^{(n_\mathrm{seq})}$ as a function of $n_\mathrm{seq}$ for two values of $\Omega$. The figure shows a clear transition from non-linear to linear behavior.

\textit{Example 3: Non-unitary dynamics.---} To demonstrate the generality of our analysis, we present that our results still hold for non-unitary dynamics. We consider the spin chain of Eq.~\eqref{eq_spin_chain} with $B{=}0$ subjected to local dissipation:
\begin{equation}
    \dot{\rho} {=}-\frac{i}{\hbar} [H, \rho]{+}\kappa\sum_{i=1}^{N}\left[(1{+}n_\mathrm{th})\mathcal{D}[\sigma^-_i]\rho{+} n_\mathrm{th}\mathcal{D}[\sigma^+_i]\rho\right],
\end{equation}
where $\mathcal{D}[O]\rho{=}O\rho O^\dagger{-}\frac{1}{2}\{O^\dagger O,\rho\}$, $\{\cdot,\cdot\}$ is the anticommutator, $\kappa$ is the decay rate, and $n_\mathrm{th}$ is the bath excitations on average. We aim to estimate $\kappa$ using local sequential measurements on the spin at site $N$. In Fig.~\ref{fig_JC_lindblad}(c), we plot the FI increment $\Delta\mathcal{F}_\kappa^{(n_\mathrm{seq})}$ as a function of $n_\mathrm{seq}$ for two values of $\kappa$. As seen from the figure, a clear constant saturation of $\Delta\mathcal{F}_\kappa^{(n_\mathrm{seq})}$ is reached for both cases, demonstrating that a finite amount of information can be extracted at each step even for non-unitary dynamics. In Fig.~\ref{fig_JC_lindblad}(d), we plot the FI $\mathcal{F}_{\kappa}^{(n_\mathrm{seq})}$ as a function of $n_\mathrm{seq}$ for two values of $\kappa$. The figure demonstrates an evident transition from non-linear to linear behavior.

\begin{figure}[t]
\includegraphics[width=\linewidth]{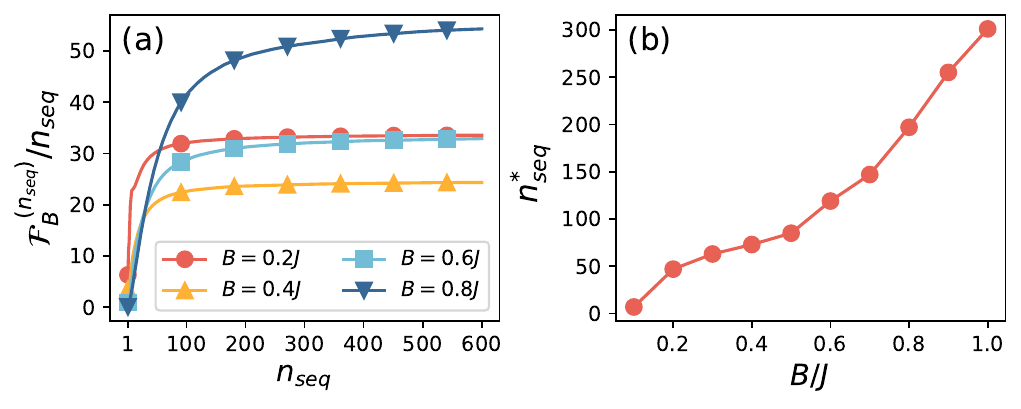} 
\caption{(a) Gain $\mathcal{F}_{B}^{(n_\mathrm{seq})}{/}n_\mathrm{seq}$ as a function of $n_\mathrm{seq}$ for different values of $B$. (b) $n_\mathrm{seq}^*$ as a function of $B$.} \label{fig_CFI_gain}
\end{figure}

\textit{Resource analysis.---} When should we reset the sensing protocol to maximize the potential of the sequential measurement scheme? To address this issue, we consider the total number of measurements, $Mn_\mathrm{seq}{=}R$, as our sensing resource $R$ (see SM~\cite{SM} for time as resource). This constraint makes the Cram\'{e}r-Rao inequality to be: $\text{Var}[\lambda]{\geq}1{/}(R\mathcal{F}_\lambda^{(n_\mathrm{seq})}{/}n_\mathrm{seq})$, where $\text{Var}[\lambda]$ is the variance of $\lambda$. Clearly, the \textit{gain} $\mathcal{F}_\lambda^{(n_\mathrm{seq})}{/}n_\mathrm{seq}$ determines the step $n_\mathrm{seq}$ to cease the protocol and initiate a new trajectory. The larger the gain, the smaller the uncertainty provided by the sequential measurement protocol. Based on the first example, in Fig.~\ref{fig_CFI_gain}(a), we plot the gain $\mathcal{F}_{B}^{(n_\mathrm{seq})}{/}n_\mathrm{seq}$ as a function of $n_\mathrm{seq}$ for different $B$. As the figure shows, the gain slows down after a specific $n_{seq}$ for all $B$. This suggests that after a certain $n_\mathrm{seq}$ the protocol provides marginal benefits. We denote $n_{seq}^*$ as the $n_\mathrm{seq}$ such that $\mathcal{F}_{B}^{(n_\mathrm{seq}^*)}{/}n_\mathrm{seq}^*$ is over $90\%$ of the saturated value at $n_\mathrm{seq}{=}600$. In Fig.~\ref{fig_CFI_gain}(b), we plot $n_\mathrm{seq}^*$ as a function of $B$. As the figure shows, $n_\mathrm{seq}^*$ grows monotonically as $B$ increases. This means that for larger $B$ one should stop the sensing protocol after longer sequential measurements.

\textit{Conclusions.---} We introduce a recursive formula and an efficient Monte-Carlo approach to evaluate the Fisher information for sequential measurements sensing of arbitrary lengths. Our findings show that the obtainable Fisher information initially grows non-linearly with respect to the number of measurements and then asymptotically saturates to a linear function. This transition is directly linked to the probe's finite memory of an early state. When the memory of such early state is lost, the information accumulation about the unknown parameter (i.e. incremental FI) becomes almost constant, resulting in linear scaling of the FI. This fundamentally limits the extractable information capacity through sequential measurements. Finally, by considering the total number of measurements as the main sensing resource, we establish a figure of merit to identify the optimal measurement sequence length. We exemplified our results in three distinct physical systems. 

\textit{Acknowledgments.---} The authors kindly acknowledge fruitful discussions with D. Burgarth, M. G. A. Paris, and J. Kahn. A.B. acknowledges support from the National Key R$\&$D Program of China (Grant No. 2018YFA0306703), the National Natural Science Foundation of China (Grants No. 12050410253, No. 92065115, and No. 12274059), and the Ministry of Science and Technology of China (Grant No. QNJ2021167001L). V.M. thanks the National Natural Science Foundation of China (Grant No. 12050410251) and the Postdoctoral Science Foundation of China (Grant No. 2022T150098)).

\bibliographystyle{apsrev4-1}
\bibliography{Seq_Meas}

%SUPPLEMENTAL MATERIAL

\clearpage
\onecolumngrid
\pagebreak
\widetext

\begin{center}
\textbf{\large Supplemental Material: Extractable Information Capacity in Sequential Measurements Metrology}

\vspace{0.25cm}

Yaoling Yang$^{1}$, Victor Montenegro$^{1}$, and Abolfazl Bayat$^{1}$

\vspace{0.25cm}

$^{1}${\small \em Institute of Fundamental and Frontier Sciences,\\ University of Electronic Science and Technology of China, Chengdu 610051, PR China}\\

\end{center}

\date{\today}
\setcounter{equation}{0}
\setcounter{figure}{0}
\setcounter{table}{0}
\setcounter{page}{1}
\makeatletter
\renewcommand{\theequation}{S\arabic{equation}}
\renewcommand{\thefigure}{S\arabic{figure}}

This Supplementary Material offers explanations on several issues, encompassing the extended proof of the recursive formula for the Fisher information, the robustness analysis of the Monte-Carlo simulation, the Jaynes-Cummings filtering achieved through repeated measurements on the qubit, and the time resource analysis for the sequential measurement sensing protocol.

\section{I. Proof of the recursive formula for Fisher information in sequential measurements metrology}

The following proof follows Ref.~\cite{zamir1998proof}. Here, we reformulate it to fit our sequential measurements sensing protocol. The Fisher information (FI) takes into account all $(n)$ sequential measurement outcomes, $\pmb{\gamma}^{(n)}{=}(\gamma_1,\gamma_2,{\cdots},\gamma_{n})$ can be expressed as follows:
\begin{equation}
    \mathcal{F}^{(n)}_\lambda=E_{\pmb{\gamma}^{(n)}}\left[\left(\frac{\partial \ln P_{\pmb{\gamma}^{(n)}}(\lambda)}{\partial \lambda}\right)^{2}\right],\label{eq_Fisher_info_expect_form}
\end{equation}
where $E[\cdot]$ is the expectation value of $\cdot$ (here, with a countable set of possible outcomes) and $P_{\pmb{\gamma}^{(n)}}$ denotes the conditional probability of the random variable $\pmb{\gamma}^{(n)}$ for $n$ consecutive outcomes.

According to the definition in Eq.~\eqref{eq:CFI_trajectories} (see main text), the probability associated with a specific trajectory exhibits the following simple relationship:
\begin{equation}
    \ln P_{\pmb{\gamma}^{(n)}}(\lambda)= \ln P_{\pmb{\gamma}^{(n-1)}}(\lambda)+\ln p(\gamma_{n} \mid \pmb{\gamma}^{(n-1)};\lambda),\label{eq_conditional_prob_sm}
\end{equation}
where $p(\gamma_{n} \mid \pmb{\gamma}^{(n-1)};\lambda)$ accounts for the conditional probability of obtaining $\gamma_{n}$ as the outcome of the $n$-th sequential measurement conditioned on $(n-1)$ previous measurements, i.e., conditioned on the trajectory $\pmb{\gamma}^{(n-1)}$.

By substituting Eq.~\eqref{eq_conditional_prob_sm} into Eq.~\eqref{eq_Fisher_info_expect_form}, one obtains:
\begin{equation}
\begin{split}
    & \mathcal{F}^{(n)} = E_{\pmb{\gamma}^{(n)}}\left[\left(\frac{\partial \ln  P_{\pmb{\gamma}^{(n-1)}}(\lambda)}{\partial \lambda} \right)^{2} \right] + E_{\pmb{\gamma}^{(n)}}\left[\left(\frac{\partial \ln p(\gamma_{n} \mid \pmb{\gamma}^{(n-1)};\lambda)}{\partial \lambda} \right)^{2} \right] \\ &+ 2\cdot E_{\pmb{\gamma}^{(n)}}\left[ \left(\frac{\partial \ln P_{\pmb{\gamma}^{(n-1)}}(\lambda)}{\partial \lambda}\right) \left(\frac{\partial \ln p(\gamma_{n} \mid \pmb{\gamma}^{(n-1)};\lambda)}{\partial \lambda} \right)\right],\label{eq_chain_rule_decomp_seq}
\end{split}
\end{equation}
where the first term on the right-hand side of Eq.~\eqref{eq_chain_rule_decomp_seq} can be written as $E_{\pmb{\gamma}^{(n)}}\left[\left(\frac{\partial \ln  P_{\pmb{\gamma}^{(n-1)}}(\lambda)}{\partial \lambda} \right)^{2} \right]=:\mathcal{F}^{(n-1)}$. The last cross term on the right-hand side of Eq.~\eqref{eq_chain_rule_decomp_seq} can be expressed as follows:
\begin{equation}
\begin{split}
    E_{\pmb{\gamma}^{(n)}} & \left[ \left(\frac{\partial \ln P_{\pmb{\gamma}^{(n-1)}}(\lambda)}{\partial \lambda}\right) \left(\frac{\partial \ln p(\gamma_{n} \mid \pmb{\gamma}^{(n-1)};\lambda)}{\partial \lambda} \right)\right] = \sum_{\pmb{\gamma}^{(n)}}  P_{\pmb{\gamma}^{(n)}}(\lambda)  \left(\frac{\partial \ln P_{\pmb{\gamma}^{(n-1)}}(\lambda)}{\partial \lambda}\right)  \left(\frac{\partial \ln p(\gamma_{n} \mid \pmb{\gamma}^{(n-1)};\lambda)}{\partial \lambda} \right) \\& =  \sum_{\pmb{\gamma}^{(n-1)}} P_{\pmb{\gamma}^{(n-1)}}(\lambda) \left(\frac{\partial \ln P_{\pmb{\gamma}^{(n-1)}}(\lambda)}{\partial \lambda}\right) \sum_{\gamma_{n}} p(\gamma_{n} \mid \pmb{\gamma}^{(n-1)};\lambda) \left(\frac{\partial \ln p(\gamma_{n} \mid \pmb{\gamma}^{(n-1)};\lambda)}{\partial \lambda} \right).
\end{split}
\end{equation}

Note that:
\begin{equation}
  \sum_{\gamma_{n}} p(\gamma_{n} \mid \pmb{\gamma}^{(n-1)};\lambda) \left( \frac{\partial \ln p(\gamma_{n} \mid \pmb{\gamma}^{(n-1)};\lambda)}{\partial \lambda} \right)
  = \partial_\lambda \sum_{\gamma_{n}} p(\gamma_{n} \mid \pmb{\gamma}^{(n-1)};\lambda)= 0,
\end{equation}

and therefore, the cross term vanishes. Eq.~\eqref{eq_chain_rule_decomp_seq} simply reduces to:
\begin{equation}
\mathcal{F}^{(n)} = \mathcal{F}^{(n-1)} + E_{\pmb{\gamma}^{(n)}}\left[\left(\frac{\partial \ln p(\gamma_{n} \mid \pmb{\gamma}^{(n-1)};\lambda)}{\partial \lambda}\right)^{2} \right].\label{eq_new_chain_rule_decomp_seq}
\end{equation}

By expanding the second term on the right-hand side of Eq.~\eqref{eq_new_chain_rule_decomp_seq}, one gets:
\begin{align}
E_{\pmb{\gamma}^{(n)}}\Biggl[\left(\frac{\partial \ln p(\gamma_{n} \mid \pmb{\gamma}^{(n-1)};\lambda)}{\partial \lambda}\right)^{2} \Biggl] \nonumber &= \sum_{\pmb{\gamma}^{(n-1)}} P_{\pmb{\gamma}^{(n-1)}} E_{\gamma_{n}}\Biggl[\left(\frac{\partial \ln p(\gamma_{n} \mid \pmb{\gamma}^{(n-1)};\lambda)}{\partial \lambda}\right)^{2} \Biggl]  \\ &=  \sum_{\pmb{\gamma}^{(n-1)}} P_{\pmb{\gamma}^{(n-1)}} f_\lambda^{\pmb{\gamma}^{(n-1)},(n)},
\end{align}
where we have defined $f_\lambda^{\pmb{\gamma}^{(n-1)},(n)}$ as the FI obtained from the $n$-th measurement probability distribution $p(\gamma_{n} \mid \pmb{\gamma}^{(n-1)};\lambda)$ in the trajectory $\pmb{\gamma}^{(n-1)}$. Therefore, the recursive formula for the FI in the sequential measurement metrology reads as:
\begin{equation}
\mathcal{F}^{(n)}_\lambda=\mathcal{F}^{(n-1)}_\lambda + \sum_{\pmb{\gamma}^{(n-1)}} P_{\pmb{\gamma}^{(n-1)}} f_\lambda^{\pmb{\gamma}^{(n-1)},(n)}.
\end{equation}

\section{II. Numerical simulations robustness analysis}
\begin{figure}[h!]
\includegraphics[width=0.95\linewidth]{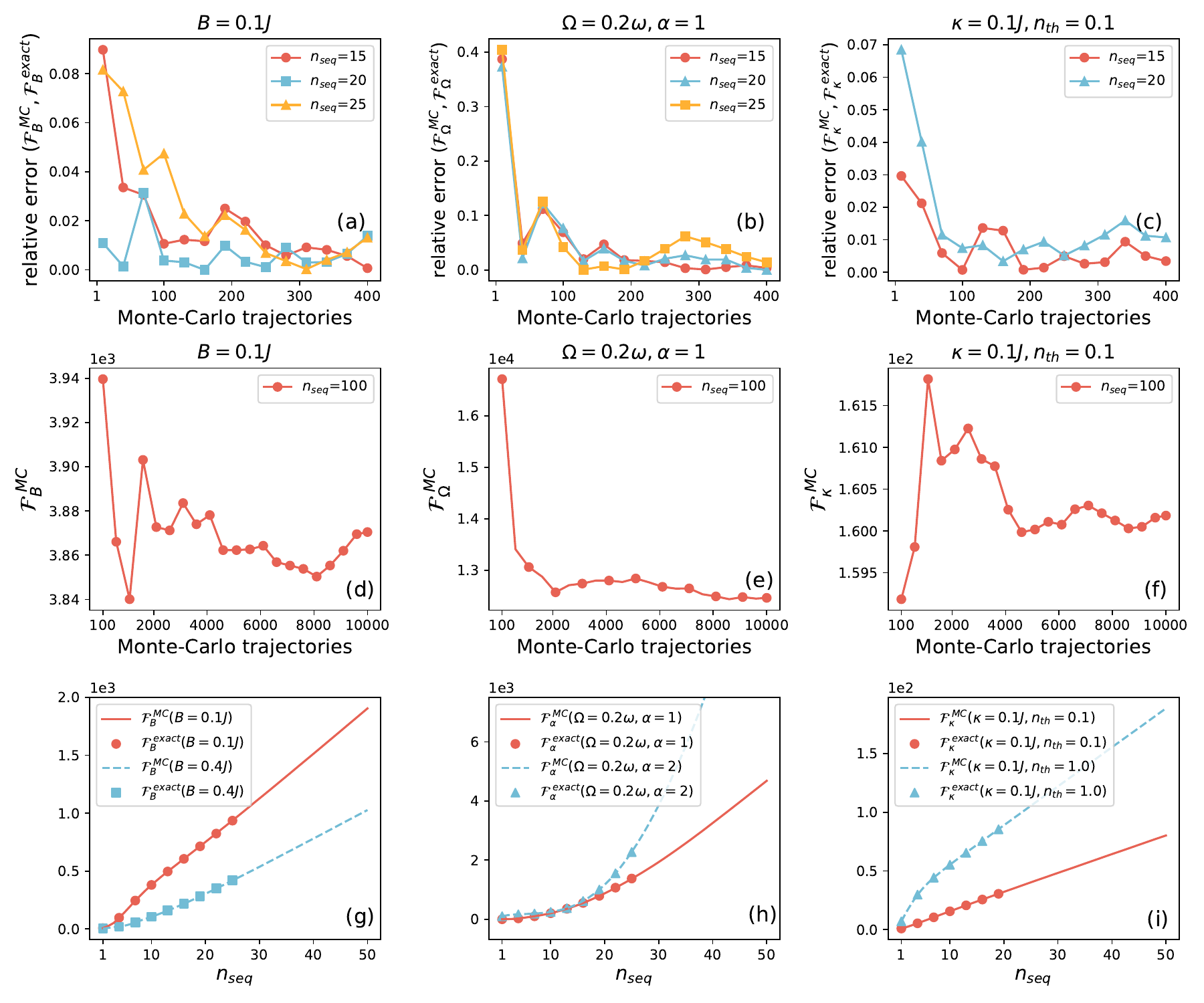}
\caption{Relative error between the Monte-Carlo FI $\mathcal{F}^{\text{MC}}$ and the exact FI $\mathcal{F}^{\text{exact}}$ as a function of the number of Monte-Carlo trajectories for (a) the spin chain, (b) light-matter interaction, and (c) non-unitary dynamics examples. The relative error decreases to around 1\% as the number of trajectories increases for all three cases. Monte-Carlo FI $\mathcal{F}^{\text{MC}}$ as a function of the number of Monte-Carlo trajectories for a fixed sequential measurement number $n_{\mathrm{seq}} = 100$, for (d) the spin chain, (e) light-matter interaction, and (f) non-unitary dynamics cases. A clear convergence of $\mathcal{F}^{\text{MC}}$ towards a stationary value emerges as the number of Monte-Carlo trajectories increases. Exact and Monte-Carlo FIs as a function of the number of sequential measurements $n_{\mathrm{seq}}$ for various parameter values, for (g) the spin chain, (h) light-matter interaction, and (i) non-unitary dynamics scenarios. The curves closely overlap, demonstrating the accuracy of our procedure.}\label{fig_SM_robustness} 
\end{figure}

This section demonstrates the high accuracy of our numerical simulations using the Monte-Carlo approximation method for calculating the FI, specifically its FI increment shown in Eq.~\eqref{eq:mc_delta_f} (see main text). Throughout this section, we use the following notation: the FI approximated using the Monte-Carlo approach is denoted as $\mathcal{F}^{\text{MC}}$, and the FI computed from exact probability distributions is denoted as $\mathcal{F}^{\text{exact}}$. Moreover, we analyze the numerical robustness for the three examples considered in our work: spin chain magnetometry, light-matter (Jaynes-Cummings) interaction, and non-unitary dynamics.

We first focus our analysis on comparing the relative error between $\mathcal{F}^{\text{MC}}$ and $\mathcal{F}^{\text{exact}}$ as the number of Monte-Carlo trajectories grows. To do so, we consider the relative error between these quantities as:
\begin{equation}
    \text{relative error}(\mathcal{F}^{\text{MC}}_i, \mathcal{F}^{\text{exact}}_i) = \frac{|\mathcal{F}^{\text{MC}}_i - \mathcal{F}^{\text{exact}}_i|}{\mathcal{F}_i^{\text{exact}}},
\end{equation}
where $i$ refers to $B, \Omega, \kappa$ for the spin chain, light-matter, and non-unitary examples, respectively. 

In Fig.~\ref{fig_SM_robustness}(a)-(c), we plot the relative error between $\mathcal{F}^{\text{MC}}$ and $\mathcal{F}^{\text{exact}}$ as a function of the number of Monte-Carlo trajectories. As the figure shows, there is a prompt reduction of the relative error for all three cases. Particularly, for this specific set of parameters, the number of trajectories needed to go below 1\% is of the order of $10^2$. It is worth noting that throughout our numerical simulations in this work, we typically used from $10^4$ to $10^5$ trajectories to ensure high accuracy as other system parameters increase, e.g., when $n_{\text{seq}}$ is of the order of $10^2$. In general, our numerical simulations are robust as the relative error reduces across all scenarios. To further support the robustness analysis, in Fig.~\ref{fig_SM_robustness}(d)-(f), we plot $\mathcal{F}^{\text{MC}}$ as a function of the Monte-Carlo trajectories for a fixed number of sequential measurements $n_{\mathrm{seq}}=100$. As seen from the figure, the FI approximated via the Monte-Carlo approach quickly converges to a stationary value across all examples. Finally, in Fig.~\ref{fig_SM_robustness}(g)-(i), we plot $\mathcal{F}^{\text{MC}}$ and $\mathcal{F}^{\text{exact}}$ individually as a function of $n_\mathrm{seq}$. It is worth noting that for the exact case, we are able to simulate up to $\sim 30$ sequential measurements, as higher values are extremely computationally costly. However, for the area in which both can be compared, the results almost overlap. Therefore, our methodology stands as reliable with very small relative error and quick convergence.

\section{III. Time as resource}
\begin{figure}[h!]
\includegraphics[width=0.95\linewidth]{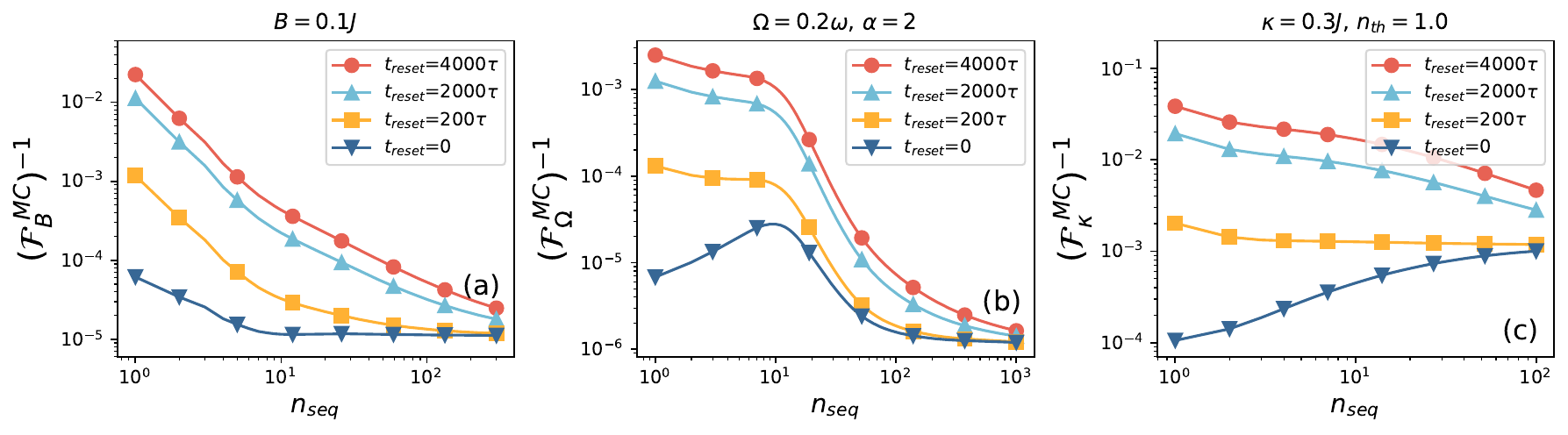} 
\caption{Inverse of the Monte-Carlo FI $\mathcal{F}^{\text{MC}}$ as a function of $n_\mathrm{seq}$ for several resetting times $t_\mathrm{reset}$: (a) Spin chain case, (b) Jaynes-Cummings case, and (c) non-unitary dynamics case. We consider the measurement time as $t_\mathrm{meas}=10\tau$, where $\tau$ is the free evolution time between measurements. In particular, we chose $J\tau=4$ for the spin chain case, $\omega\tau=2\pi$ for the Jaynes-Cummings case, and $J\tau=1$ for the non-unitary case, respectively. Other parameters are shown in the figure.} \label{fig_time_as_resource} 
\end{figure}

In the main text, we consider the total number of measurements as our main sensing resource. With this approach, we were able to determine the specific $n_\mathrm{seq}^*$ at which we needed to stop that trajectory and start a newly fresh trajectory. This made us to exploit the sequential measurements scheme optimally with respect to that sensing resource. Nonetheless, time can also be accounted for as a sensing resource. To consider time as a resource, we follow Ref.~\cite{PhysRevLett.129.120503}, where the total protocol time $T$ is defined as:
\begin{equation}
T = M(t_\mathrm{reset} + n_\mathrm{seq}t_\mathrm{meas} + n_\mathrm{seq}\tau),
\end{equation}
where $M$ is the total number of trajectories, $t_\mathrm{reset}$ is the time it takes to reset each trajectory, $t_\mathrm{meas}$ is the time it takes to measure the particle, $\tau$ is the free evolution time between measurements, and $n_\mathrm{seq}$ is the number of sequential measurements. It is typical in experiments that the resetting time is larger than the time it takes to measure a subsystem. Thus, we consider $t_\mathrm{meas}=10\tau$, while $t_\mathrm{reset}$ is found to be between $0\leq t_\mathrm{reset}\leq 
 4000\tau$.

By recalling the Cram\'{e}r-Rao inequality, one obtains that the variance of the unknown parameter relates to the inverse of the FI, i.e., $\mathrm{Var}[\lambda] \geq (\mathcal{F})^{-1}$. Therefore, the lower the FI inverse, the lower the uncertainty.

Having this relationship at hand, in Fig.~\ref{fig_time_as_resource}(a)-(c), we plot the inverse FI (approximated using the Monte-Carlo approach) as a function of $n_\mathrm{seq}$ for several values of $t_\mathrm{reset}$ for a fixed total protocol time $T$. As seen from the figure, all cases show that the best scenario is the \textit{ideal} case where $t_\mathrm{reset}=0$. However, once $t_\mathrm{reset}\neq 0$, the results show that one can truly benefit from consecutively measuring the system. Interestingly, all panels show that for the same fixed total protocol time $T$, a very long number of sequential measurements proves to be very beneficial for reducing the uncertainty of the unknown parameter. In other words, since resetting the system is very costly, one could spend the entire time measuring the system constrained to the same total protocol time $T$, achieving similar sensing performance with a very long number of sequential measurements.

\section{IV. Jaynes-Cummings filtering towards a Fock number state}
\begin{figure}[h!]
\includegraphics[width=0.95\linewidth, left]{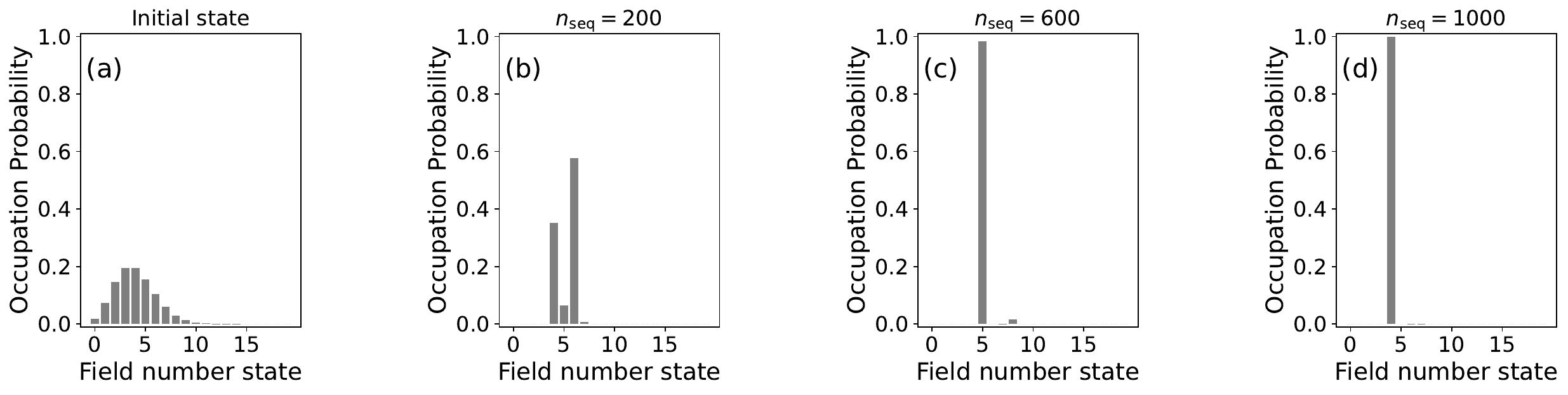}

\vspace{1cm}

\centering\includegraphics[width=\linewidth]{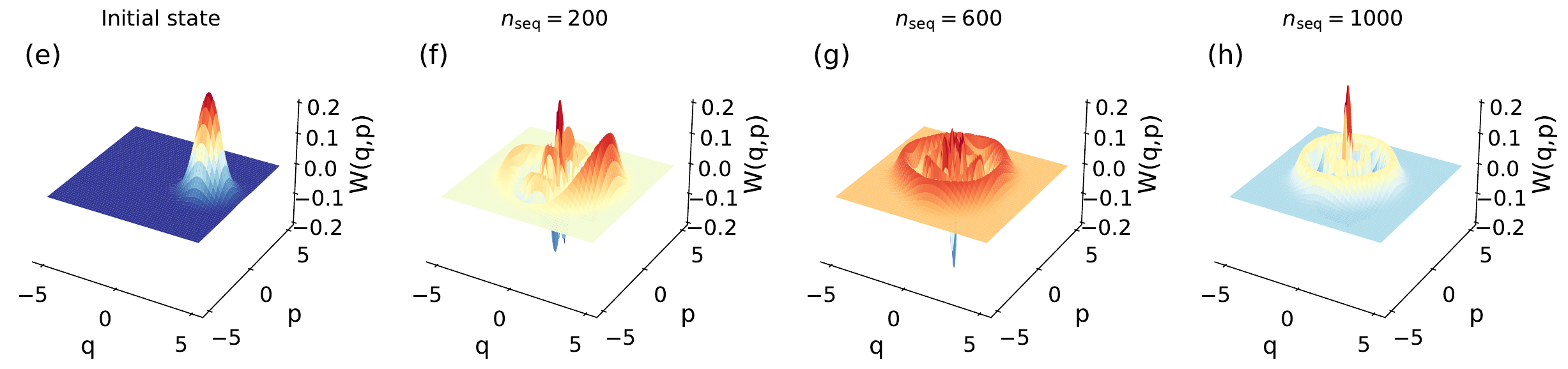}
\caption{The top row, panels (a) to (d), shows the occupation probability of the field as a function of its number state for different $n_\mathrm{seq}$. The bottom row, panels (e) to (h), illustrates the Wigner function for the field state for different $n_\mathrm{seq}$ instances. This figure demonstrates that the field, under Jaynes-Cummings interaction, is likely to collapse into a single number state after the qubit is measured repeatedly.} \label{fig_SM_JC} 
\end{figure}

For the Jaynes-Cummings example, we stated that after many sequential steps, the cavity field is likely to collapse into a single number state, denoted $|\tilde{m}\rangle$ in the main text. Here, we present numerical evidence to support such a statement. 

Let us start from the Jaynes-Cummings Hamiltonian  $H_\mathrm{JC}$ [see main text Eq.~\eqref{eq_JC_model} for details on the notation]:
\begin{equation}
H_\mathrm{JC}=\hbar\omega a^\dagger a+\frac{1}{2}\hbar\omega \sigma^z+\hbar\Omega(\sigma^+a+\sigma^-a^\dagger).
\end{equation}
By initializing each trajectory from $|\psi(0)\rangle=|g\rangle|\alpha\rangle$, where $|\alpha\rangle=\sum_m\mathcal{C}(m)|m\rangle$ is a coherent state of amplitude $\alpha$, here $\alpha\in\mathbb{Re}$, and
\begin{equation}
\mathcal{C}(m)=e^{-\frac{\alpha^2}{2}}\frac{\alpha^m}{\sqrt{m!}},
\end{equation}
one obtains the evolved wave function as:
\begin{equation}
|\psi(t)\rangle=e^{-i\omega t H_\mathrm{JC}}|\psi(0)\rangle=\sum_m \mathcal{C}_g(m)|g,m\rangle+\mathcal{C}_e(m)|e,m-1\rangle,
\end{equation}
where the field distributions associated with the states $|g\rangle$ and $|e\rangle$ are:
\begin{eqnarray}
\mathcal{C}_g(m)&=&\mathcal{C}(m)e^{-im\omega t}\cos(\sqrt{m}\Omega t),\\
\mathcal{C}_e(m)&=&-i\mathcal{C}(m)e^{-im\omega t}\sin(\sqrt{m}\Omega t).
\end{eqnarray}

Note that the initial probability field distribution, namely $\mathcal{P}_0(n)=|\mathcal{C}(n)|^2$, is now \textit{split} into $\mathcal{P}_g(n)=|\mathcal{C}_g(n)|^2$ and $\mathcal{P}_e(n)=|\mathcal{C}e(n)|^2$. This is the core of the filtering process. By performing local measurements $n_\mathrm{seq}$ times on the qubit sequentially, here at (scaled) times $2\pi$, the number state distribution filters to new distributions associated with states $|g\rangle$ or $|e\rangle$.

In Fig.~\ref{fig_SM_JC} we present the sequential measurements procedure for a representative trajectory. The first row of Fig.~\ref{fig_SM_JC} [panels (a) to (d)] shows the field probability occupation as a function of the field number. As the figure shows, in Fig.~\ref{fig_SM_JC}(a) the initial coherent distribution is subsequently filtered as the number of $n_\mathrm{seq}$ increases. Collapsing ultimately into a single field state $|\tilde{m}\rangle$ as seen in Fig.~\ref{fig_SM_JC}(d). Note that, this represents a likely trajectory to happen, however, other field distributions can indeed occur. This is because the distributions accompanying the qubit states $\mathcal{C}_g(m)$ and $\mathcal{C}_e(m)$, see above discussion, depend on $\Omega$, the number state $m$, and the measurement time $t$ as well. Therefore, the filtering is conditioned upon other system's parameters. Unwanted collapsing states can further be filtered by properly tuning the above. To further discussing the filtering case, we consider the Wigner quasi-probability distribution of the field to observe its behaviour in phase-space. The Wigner quasi-probability distribution is defined as following:
\begin{equation}
W(q, p) = \frac{1}{\pi \hbar} \int_{-\infty}^{\infty} \langle q - y | \rho_\mathrm{field} | q + y \rangle e^{2ipy/\hbar} dy.
\end{equation}
In the bottom row of Fig.~\ref{fig_SM_JC}, we show the Wigner function $W(q,p)$ for different $n_\mathrm{seq}$, namely collapsed instances of the field state. As the figure shows, the initial field distribution shown in Fig.~\ref{fig_SM_JC}(e), $W(q,p)>0$, becomes more and more as $n_\mathrm{seq}$ increases. The final panel depicted in Fig.~\ref{fig_SM_JC}(h), shown a clear single field number state $|\tilde{m}\rangle = 4$.

\end{document}